\documentclass[12pt,letterpaper]{article}
\usepackage{changes}

\usepackage[utf8]{inputenc}

\usepackage{fixltx2e}

\usepackage{amsmath,amssymb}

\usepackage{cite}

\usepackage{nameref,hyperref}

\usepackage[right]{lineno}

\usepackage{rotating}

\makeatletter
\renewcommand{\@biblabel}[1]{\quad#1.}
\makeatother

\date{}

\usepackage{lastpage,fancyhdr,graphicx}
\usepackage{epstopdf}

\begin{document}
\vspace*{0.35in}

\begin{flushleft}
{\Large \textbf\newline{Single gene dynamics under controlled
mating} }
\newline
\\
Vladimir Obolonkin\textsuperscript{1,*}, Anatoly
Zherelo\textsuperscript{2}, George Krylov\textsuperscript{3},
Murray Jorgensen\textsuperscript{4}, Richard Spelman\textsuperscript{1}
\\
\bf{1} Livestock Improvement Corporation, Hamilton, New Zealand
\\
\bf{2} Department of Nonlinear and Stochastic  Analysis, Institute
of Mathematics, National Academy of Sciences, Minsk, Belarus
\\
\bf{3} Department of Physics, Belarusian State University, Minsk,
Belarus
\\
\bf{4} Department of Statistics, University of Waikato, Hamilton,
New Zealand
\\
%
%

* E-mail: vobolonkin@lic.co.nz
\end{flushleft}
\section*{Abstract}
We seek models for the genotype evolution of agricultural animals,
animals involved in primary production processes. Classical models
for genotype evolution have tended to be very simple in order that
analytic methods may be employed in their study. Unfortunately
these models fail to describe processes in artificially controlled
populations including agricultural livestock. It is particularly
important {to describe such processes} in order to make better use
of {the} massive genotyping data becoming available.

We describe an approach to stochastically modeling the {dynamics}
of a biallelic polymorphism {herds}
 under conditions of controlled mating {and restriction  of herds size from above}. The system of
stochastic differential equations that we propose is based on jump
diffusion processes to provide an effective platform for Monte
Carlo simulation. Our choice of this modeling framework
foreshadows the use of semi-analytic tools to complement
simulation. {Another reason for adopting the framework is its
flexibility in modeling different population management systems.}

A feature of the model is the division of the  {population} into a
{\it main herd} comprised of animals involved in the production
process and a {\it replacement herd} of {animals not currently in
the production process, typically juvenile animals}. This feature
allows for exploring different strategies for adding replacement
animals to the main herd without altering the part of the model
concerned with the dynamics of the main herd.

A discrete-time version of the model has been developed which
reflects the typical practice of New Zealand dairy herd
management.

Our Monte Carlo simulation has demonstrated that an isolated deme
whose size is bounded above (by imposition of a fixed size control
requirement) demonstrate size stabilization at a level less than
the control limit, it is looks like partial extinction, the effect
being well known in classic models.  {Another interesting feature
of the model with a size control rule is its sensitivity to a form
of a control. We have found that even change a rule to different
moment of choice of animal substitution ( from replacement herd to
a main one) results in observable variation in herds' temporal
characteristics.} {We demonstrate several simulation results under
the condition of Mendelian inheritance and its corresponding rule
of  summation. We also propose a variant of the model taking into
account  animal inflows and outflows providing exchange through an
external market.}  For future work we consider the cooperative
development of an open source platform for such modeling and for
{\it in silico} experiments utilizing real genotyping data from
the New Zealand dairy cow population.


\section*{Introduction}
The two seminal papers of G. Hardy \cite{Hardy} and W. Weinberg
\cite{Weinberg} on the steady state distribution of alleles were
based on the Mendelian law of genetics. The Mendelian law and
{these papers} are the corner stones of practical genetics and
strongly influenced later development in the field.

One of the main results, called {\it Hardy-Weinberg equilibrium},
states that in a population {satisfying certain conditions} the
observed frequencies of possible genotypes $AA, AB, BB$ in some
locus of interest are $p^2, 2pq, q^2$,  where $p$ and $q=1-p$ are
the proportions of the alleles $A$ and $B$ at the locus. In
reality, the conditions needed to ensure Hardy-Weinberg
equilibrium often fail to be met. For example in the case of New
Zealand dairy cows, the national herd  has about
$4\times 10^6$ cows distributed in about $1.1\times 10^4$ herds. 
There were just over 3700 bulls used for insemination in 2013-2014
season with under 100 top bulls used to mate 80\% of the whole
national herd. Except by chance the genotype frequencies in whole
population are far from Hardy-Weinberg proportions at any locus
and the proportions vary by herd, region, and loci.
Table~\ref{hwtab} compares the conditions needed to establish
Hardy-Weinberg equilibrium with the production situation in the
dairy industry in New Zealand and, we suspect, many other
countries.

In {reality} such steady-state results are not of central interest
for animal breeding decisions as the intent is to change the
properties of the production herd in a favored direction. Before
considering a more realistic model for such controlled breeding we
will set the scene by mentioning some classical dynamic models in
genetics.

One of the most influential such models was a genetic evolution
model proposed by S. Wright  \cite{Wright} and R. Fisher
\cite{Fisher}. In this model, for a fixed size total population
and binary alleles ($A,B$),
  the discrete time dynamics of relative frequencies of
different types of individual is considered under the neutrality
({\it equal fitness}) and Markov assumptions. This allows a
stochastic dynamic flow with rather good analytical properties.
The model has the same probability structure that would result if
{\em each of the $N$ individuals of the $n + 1^{\mbox{st}}$
generation picked their parents at random}, though of course this
cannot literally happen. A prominent phenomenon of the neutral
Wright-Fisher model without mutation is a ``fixation'' that is the
extinction of all
 types but one at a finite but random time.

The further development of the Fisher approach has been in  ways
of weakening the assumptions used. The two most evident ones are
to introduce more types of alleles \cite{Feng} and to consider a
variable resampling rate (floating total population size)
\cite{Donnelly, Kaj-and-Krone}.
More sophisticated approaches are to introduce a more complicated distribution law (%
say, a Poissonian one) for a number of offspring of an individual
\cite{Feng}, to take into account mutation and/or selection
\cite{Steinsaltz}, to introduce a random process for mortality of
individuals \cite{Moran}, or to work with diffusion-like models
\cite{Feller}. Good introductions to some of these general topics
can be found in \cite{Dawson, PPH} and the first chapter of
\cite{Feng}.

These models  are relatively simple in that analytical results for
variables of interest, or asymptotic expansions for them, may be
derived within them (see for example \cite{DPP}). But to make the
obtaining of analytical results a goal inevitably leads to a focus
on mathematical tractability in the setting up of the model. This
in turn encourages the avoidance of the complexities of real life
problems, resulting in an oversimplified model.

An important point is that classical models typically assume {\it
infinitely large} population size while investigating the dynamics
of allele frequencies. While these models have powerful asymptotic
analysis techniques available to them they are not relevant to the
situation of a typical farm running a herd of only a few hundred
cows. So instead we will concentrate on accounting for the effects
of a finite, stochastically driven, herd size.

While in this paper we concentrate on cows on a particular farm it
is important to note that these farm herds are strongly connected
by the fact mentioned above that a few tens of top bulls fertilize
millions of New Zealand cows. This focus on a small number of male
animals is in further contrast to the situation envisioned by the
classical population models.

Another difficulty with the quest for analytic solutions is that
often they are available only for the evolution of mean values of
variables of interest and not for other properties of their
distributions such as variance and shape, whose knowledge will be
required in real industry applications. Thus it is necessary to
put analytic methods to the side and develop realistic stochastic
models allowing Monte-Carlo simulation of all required
distributions.

The chief tool available to the dairy industry for improvement of
the genetic merit of the herd is the ability, via artificial
insemination, to choose the parents of the next generation. So our
concern is chiefly with genotype dynamics under {\it controlled
mating}, often termed {\it artificial breeding} in the dairy
industry. Genomic methods are becoming important in sire selection
now that statistical genomic prediction methods are available
\cite{Harris}.  Hand in hand with this animal {evaluation}
technology vast genotyping data sets are increasingly available.

The model we develop in this paper is a rather general one in
sense  of being capable of a range of adjustments. For example
other species could be considered, as might
more sophisticated rules of genotype sum 
possibly addressing genotype dependent ({\it genomic}) selection,
or other departures from the {\it equal fitness} assumption. {Our
model simulates the dynamics of alleles at a single locus in each
animal of a herd under controlled mating.}

The model, in the discrete form given below, incorporates some
features typical for New Zealand seasonal dairy herd management
practice. This allows the model to be useful for estimating the
time necessary to reduce the proportion of unwanted genomic
variants in the population to an acceptably low value. Also it
could be used to estimate the time to introduce some desirable
genomic variant, for example one that influenced milk composition
in a beneficial way.

There are a growing number of genomic discoveries published
concerning the importance of some particular single variants
\cite{BernardGrisart}, \cite{Aurelie}, \cite{Morris}. It is
obvious the number of known deleterious genes will increase as
knowledge is gained and more is understood about common diseases.
Every sire selected for AI carries some deleterious genes. The
model developed  may help in risk analysis by running different
scenarios to optimize AI strategy in sense of  performance merit
vs deleterious gene carrying.

\section*{Methods}
Firstly we introduce a general stochastic model for genotype
evolution in an isolated herd subjected to a maximum herd size.
Secondly we present a discrete-time version of this model as well
as a generalization which allows a limited inflow of animals from
an external source such as the market. The discrete model is then
used in few simulations to illustrate its use, obtaining some
interesting results. Finally we outline directions for possible
further development.

Let us consider a herd as effectively comprising two sub-herds:
the main ({\it production}) herd and a {\it replacement} herd. The
 main herd consists of adult animals providing the productive output of a herd e.g. cows in milk.
The replacement herd includes mostly young animals from birth up
to just before going into production. It can also include some
(typically small) number of adult animals, each expected to be
suitable when required to join the main herd as a replacement.

We will  {(for example in equations \ref{eqn-main} below)} use
subscripts $i$ and $j$ to refer to individual members of each herd
but these numbers will refer to a formal position in a herd, much
as do the numbers on the shirts of football players in a team. We
will use functions of the form $f_i(t)$ to refer to properties of
the animal in position $i$ of the herd as a function of time, $t$.
If a maximum herd size is imposed, a new animal can only be
introduced as a replacement for a removed animal or if the herd
size is below the limit. In the latter case it will take the first
unoccupied position. If animal $i$ is replaced at time $t_*$ this
will typically cause a discontinuity or `jump' in the function
$f_i(t)$ at $t=t_*$.

The presence of animal replacements means the stochastic processes
in our model will have jumps. Such processes have found much
application in financial modeling. It has been found that the best
way to formulate these processes is to use the integral form of
 {stochastic differential equations} using  {stochastic}
Ito integration (\cite{Oksendal,Applebaum}). Jumps usually arise
in financial modeling as the result of a real-world event changing
the value of a stock. The jumps in our model will not be of this
kind. An analog of our type of model in Finance might be where a
portfolio of stocks corresponds to a herd of animals and a jump is
caused by replacing one stock in the portfolio by a new stock.

Our model {makes} a number of assumptions about {the main and
replacement} herds and some parameters of animal movements in and
out of the herd.
\subsection*{List of assumptions for continuous time model}
\begin{enumerate}
\item The number of cows in the main herd is initialised to $N_0$  at the initial time $t_{0}=0$ and never exceeds this value subsequently.
The corresponding dynamical variable is $N(t)$.
\item The number of cows in the replacement herd is initialised to $M_0$  at the initial time $t_{0}=0$ and never exceeds this value subsequently.
The corresponding dynamical variable is $M(t)$.

\item At initial time $t_{0}=0$, the age of cows  in the main herd is generated by a customized random number generator.
\item The genotype of a progeny follows from that of its parents
via a summation rule. In this article we consider
 the Mendelian only case expressed by
 equation \ref{eq:sum-rule}.
\item\label{item-depart} The departure of an animal from the main herd is subject to a Poisson process with a rate parameter $\lambda_D$.
This allows  uniform accounting for different causes of animal
departure ({\it animal fate}).
\item\label{item-departure} The departure of an animal from the replacement herd is also subject to a Poisson process with another rate parameter $\lambda_d$.
\item\label{item-movement} The animal movements between sub-herds and departures happen annually (once a year) and are simulated by the following scheme.
Using assumptions 1 to 6 simulate this year's set of animals to
depart. Then fill vacancies thus created in the main herd by
random choice  (variable $\xi$) from members of the replacement
herd that have reached the age of $t_{min}$ years to maintain
predefined size. If this turns to be impossible due to lack of
heifers of proper age in the replacement herd, then replace as
many as possible, the main herd now taking a smaller size.

\item The replacement strategy in the replacement herd is annual (once a year) addition of  newborn calves
to maintain predefined size $M$. If it turns to be impossible due
to lack of newborn calves, the replacement herd remains with this
smaller size.
\item One bull, or team of bulls, of known genotype sires the whole herd. A new generation appears every year.
\item There is no in-flow of animals from outside.
\end{enumerate}

{Assumptions \ref{item-depart} and \ref{item-departure} are for
simplicity and may later be replaced by other descriptions of
these animal departure processes more closely reflecting actual
herd management practice.}

Assumption \ref{item-movement} {somewhat departs from the} common
practice, which is to have some flow of animals from outside (say,
from the market) but we leave this for a subsequent publication
considering the modeling of multiple herds. There is some
discussion of market influence below; see {\it Herd with a
limited  inflow of animals}.

The parameters $\lambda_D, \lambda_d$ are chosen by estimating the
mean  of animal's life time in the main and replacement herds
respectively either on common practice or detailed analysis of the
survival curve.  {In New Zealand practice actual mortality as a
cause of departure from either herd would be rare but poor
condition might cause a decision to remove an animal from either
herd}

Variants of the model with different strategy, distribution and
parameter settings are possible but {are outside }
the scope of this article.

\subsection*{Rule of single genotype sum}\label{sec-rule-of-sum}

We represent the genotype of an animal in the locus of interest as
a number from $\{-1,0,1\}$. Where $-1$ and $1$ stand for
homozygous and $0$  for heterozygous genotypes correspondingly.

The mode of inheritance at a single locus assumed to be {\it
Mendelian} leading to the following {\it rule of summation} (where
$P$ gives the probability of each outcome):

\begin{eqnarray} \label{eq:sum-rule}
(-1)\dot +(-1)&=&-1,\ \ P=1 \nonumber \\
0\dot +(-1)&=&\begin{cases}
-1&P=0.5\\
0&P=0.5
\end{cases} \nonumber \\
1\dot +(-1)&=&0,\ \ P=1 \\
0\dot +0&=&\begin{cases}
         -1, & P=0.25\\
         0,  & P=0.5\\
         1,  & P=0.25
        \end{cases} \nonumber \\
 1\dot +0&=&\begin{cases}
         0, & P=0.5\\
         1, & P=0.5
        \end{cases}  \nonumber \\
 1\dot +1 &=&1,\ \ P=1 \nonumber
\end{eqnarray}

Here $\dot +$ is a commutative infix operation giving a random
value of a `child' genotype as a random function of two variables
of  corresponding parental genotypes.

\subsection*{Continuous time model: integral form}
 {Our goal in this article has been to introduce a very
general model with the flexibility to represent a variety of types
of managed animal populations.} To express  {such a}
 model in the language of Stochastic Differential Equations and so have access to that body of theory we need to represent time in a continuous manner. Models with continuous time can show closeness to the observable
herd dynamics but require some modification to assumptions
\ref{item-departure} and \ref{item-movement}
in 
the list of model assumptions given above regarding the random
jump time for the departure processes.

 We will use the index $j$ for the above-mentioned formal position in a main herd and the
 index $i$ for the same in the replacement one.
 To develop our stochastic model of two interacting herds we state two elementary evolution
 processes for every index.  The first is the process of the changing genotype value in a position ($j$ or $i$), designated as $D_j(t)$ or $d_i(t)$. The second is the process of the changing animal
 age in a position, designated as $A_j$ and $a_i$. These  processes are responsible
 for decisions on position characteristics such as whether or not to replace an animal.

 The control on animal departure from the herds will be based on additional
 independent processes $P_{D_j}$ and $P_{d_i}$ in such a way that changes of $D_j(t)$ and (or) $d_i(t)$
 occur precisely at the time moment of a jump of the appropriate Poisson process $P$.

Accounting for this and based on the previous discussion, we can
write down a system of stochastic  equations describing temporal
evolution of ensemble of cows in main and replacement herds in the
following form

\begin{equation}
\begin{array}{ll}\label{eqn-main}
    D_j(t) &=\ D_j(0) + \int_0^t \left( -D_j(s-) + d_{\xi_j} (s-) \right) d P_{Dj} (s+A_j(0)) \\
    d_i(t) &=\ d_i(0) + \int_0^t \left( -d_i(s-) + f(D_{\eta_i} (s-), S_{\zeta_i}) \right) d P_{d_{i}} (s+a_i(0)) \\
    A_j(t) &=\ A_j(0) + t - \int_0^t ( A_j(s-) - a_{\xi_j}(s-) ) dP_{Dj}(s+A_j(0)) \\
    a_i(t) &=\ a_i(0) + t -  \int_0^t a_i(s-) dP_{d_{i}}(s+a_i(0))\\
 \end{array}
\end{equation}

 where $t\in [0,T]$, and the other quantities in equation~(\ref{eqn-main}) are defined as follows

 \begin{itemize}

  \item $D_j$ is the value of the allele for a formal $j$-th cow in the main herd;

  \item $P_{D_j}$ is the Poisson process with a parameter
        $\lambda_D$ defining the elimination rate for a cow in the main herd;
  \item $A_j$ is the age  of the $j$-th cow of the main herd;
  \item $d_i$ the value of the allele for a formal $i$-th cow of the replacement herd;
  \item $P_{d_i}$ is the Poisson process with a parameter
        $\lambda_d$ defining the elimination rate for a cow in the replacement herd;
  \item $a_i$  is the age of the $i$-th cow of the replacement herd;
  \item $f(\cdot,\cdot)$ is the female calf's genotype as a stochastic function of the parental genotypes. Here $f(\cdot,\cdot)$ is usually given by the rule of summation, equation~\ref{eq:sum-rule}, but other rules are possible.
  \item $\eta$ is a random variable corresponding to the random choice
        of a cow from the main herd to be used as a dam for the replacement
        herd and which subsequently gives birth to a female animal.

  \item $\{S_k\}$ is a set of values of alleles for sires;
  \item $\zeta$ is a rule for choosing the sire; it can be a random variable or a determined sequence.

  \end{itemize}

We explain the correctness of the system (2) for a position $j$ in
the main herd as follows.  A jump of the Poisson process for the
first equation occurs at a moment $s$. Due to the definition  of
the Ito integral the new value of the locus is formed by
arithmetic summation with terms $-D_j(s-)+d_\xi(s-)$. The first
term zeros the current position value and the second one
establishes the new value (with a random choice $\xi$). Based on
the same Poisson process jump the third equation in (2) reflects
the change at position $j$ of the age variable $A_j$, taking into
account that age should be increased  if there is no jump (term
$t$ in this equation).  In a similar way we deal with replacement
herd except for the fact that the new genotype value at the locus
is defined by the summation rule (1). The connection of the
replacement herd with a main one is fulfilled via the term
$D_\eta(s-)$ as an argument in function $f$ defined above.

At a given stage of the model construction we assume that all
variables and processes are mutually independent{, for example
that the loss of a cow from the main herd is not affected by
losses in the replacement herd.} Integrals are in Ito's sense, see
for example \cite{Gardiner}, p. 84.

The definition of integrals over a Poisson process can be done by
taking into account the fact that $P(t) = \widetilde{P}(t)+\lambda
t$, where $P(t)$ is the Poisson process with a parameter
$\lambda$, and  $\widetilde{P}(t)$ is a martingale corresponding
to $P(t)$ known as the {\it compensated Poisson process}.

An important point to be mentioned is that the proposed model is a
rigorously defined system of stochastic differential equations in
integral form. One possible alternative approach to the modeling
would be to proceed as is done in Evolutionary Game Theory
\cite{MSmith} where one  defines a stochastic dynamical flow by
set of local ``game rules'', an approach which is also suitable
for Monte-Carlo simulation of system evolution. Formulation in the
form of stochastic differential equations also allows Monte-Carlo
simulation but does not restrict itself to this. For example
approximate methods of the so-called weak type
\cite{Egorov,Zherelo} exist which allow (admittedly in a rather
difficult way) direct  estimates of functionals constructed on
solutions of stochastic differential equations, such as the
functionals for mean and variance of variables of interest. These
direct estimates do not require exhaustive Monte-Carlo simulation
to ensure proper statistic quality of simulation results.

Leaving this possibility for future work, we turn to discrete-time
methods. So now let us reformulate the model in the discrete form
suitable for Monte-Carlo simulation.

\subsection*{The model in discrete form }\label{sec-discrete}

{Digital computers operate in a discrete world and so for
simulation purposes, it is convenient to work with a model in
discrete time. Note also that we would commonly lack precise
information on the timing of events in a herd but may have this
information on a monthly or annual basis. We} discretize the above
model at a sequence of fixed time steps
 $0 = \tau_0 < \tau_1 < \ldots < \tau_L = T$ {}.

{
 We suppose that in each of the $L$ intervals $(\tau_{l-1}, \tau_l]$  the probability of more than one
 Poisson process jump (control action) within the interval is negligible, or alternatively that multiple jumps can be replaced by a single jump of value equal to the sum of the individual jumps.
}

 Then we
arrive at the following discrete system of model equations:

\begin{equation}
\begin{array}{ll}\label{eqn-discrete}
     D_j(t) &= D_j(0) + \sum_{l=1}^L (-D_j(\tau_{l-1})+ d_\xi(\tau_{l-1}))
      \mathrm{sign} [ \mathcal{P}(\lambda_D A_j(\tau_{l-1})) ] I_{[0,t )}(\tau_{l}) \\
  d_i(t) &= d_i(0) + \sum_{l=1}^L \left(-d_i(\tau_{l-1})+ f(D_\eta(\tau_{l-1}), S_\zeta)
      \right)\mathrm{sign} [ \mathcal{P}(\lambda_d a_i(\tau_{l-1})) ] I_{[0,t )}(\tau_{l}) \\
  A_j(t) &= A_j(0) + \sum_{l=1}^L (1 + (a_\xi(\tau_{l-1}) - A_j(\tau_{l-1}))
  \mathrm{sign} [ \mathcal{P}(\lambda_D A_j(\tau_{l-1})) ])I_{[0,t )}(\tau_{l}) \\
  a_i(t) &= a_i(0) + \sum_{l=1}^L (1 - a_i(\tau_{l-1})
  \mathrm{sign} [ \mathcal{P}(\lambda_d a_i(\tau_{l-1})) ])I_{[0,t
  )}(\tau_{l}).\\
 \end{array}
\end{equation}

In the system of equations~(\ref{eqn-discrete}), the
$\mathcal{P}(\kappa)$ are all independent Poisson random values
with rate parameters

\subsection*{Herd with a limited  inflow of animals}\label{sec-market}

The model described in the previous section for a medium herd size
can demonstrate a long period of animal deficit due to {an
``extinction effect'' which has been mentioned in the literature
and which is} observable in our simulation study. (see the
discussion of Figure~\ref{fig:4ab} below). Such a situation seems
not to be a typical one as a farmer tends to fill the gap by
animal purchase.

To account for such {herd size control polices}, we will modify
the model in the following way. We assume that a purchased animal
is placed into the replacement herd first. Such a simple
assumption, nevertheless allows us to incorporate a market inflow
by reformulating the meaning of the $\eta$ variable in
equation~(\ref{eqn-discrete})  only. Namely, {a} a zero value of
this variable now will correspond to the {\em event of a male
animal birth} whereas for born female the variable value is still
the index of the dam in the main herd. Then we can rewrite the
equations for the replacement herd in the following simple form

\begin{align} \label{eqn-discrete-market}
  d_i(t) =& d_i(0) + \sum_{l=1}^L \big(-d_i(\tau_{l-1}) 
      +\mathrm{sign}(\eta) f(D_\eta(\tau_{l-1}), S_\zeta) \nonumber\\
      & + (1-\mathrm{sign}(\eta)) D_M \big)\mathrm{sign} [ P_d(\lambda_d a_i(\tau_{l-1})) ] I_{[0,t )}(\tau_{l}) \nonumber \\
  a_i(t) =& a_i(0) + \sum_{l=1}^L \bigg(1 +\big( (1-\mathrm{sign}(\eta) )a_M  \nonumber \\
      & - \mathrm{sign}(\eta) a_i(\tau_{l-1})\big) \mathrm{sign} [ P_d(\lambda_d a_i(\tau_{l-1})) ] \bigg) I_{[0,t)}(\tau_{l}).
\end{align}

where $D_M$ is the random variable defining the distribution of
the modeled allele in animals from the market, $a_M$ is the random
variable for the cows age distribution at the market.

As one can see from the equations~(\ref{eqn-discrete-market}),
when $\eta=0$ (male is born) the $\mathrm{sign}$-function gives a
non-zero contribution into terms with index $M$ and the last can
be interpreted as incorporation of animal from market into the
replacement herd. When $\eta>0$, the factor
$(1-\mathrm{sign}(\eta))$ zeros the market contribution, restoring
the original system (\ref{eqn-discrete}).

One additional technical advantage of the last proposed model is
that now the  size of the replacement  herd is constant, which
simplifies working with the system, especially for the goals of
numerical simulation.

\section*{Simulation of genotype dynamics}\label{sec-sim}

{In order to account for genomic selection models in animal
breeding we} now show some simulation results under the rule of
genotype sum and controlled mating. We continue using the $-1, 0,
1$ coding. In each example we begin with a dam of genotype $-1$
(homozygous with allele to be eliminated), the dam and her
resulting progenies are then inseminated by a sequence of sires of
known genotypes. {We choose the time points for the discretization
with a constant one year spacing, that is we set  $\tau_l -
\tau_{l-1}=1$ (year), $l=0,1,\ldots,L$.} At the initial time
{$\tau_0$} it is also assumed that the distribution of alleles in
the replacement herd is the same as in the main one. This leads,
as we see later, to a two-year lag in switching dependence.

First, we consider the dynamics for unconditional switch of
genotype into state $1$ (homozygous with allele to be introduced).
We achieve the goal by the sequence of sires, where all sires have
genotype $1$.

\begin{figure}[h]
\begin{center}
\includegraphics[width=5.5in]{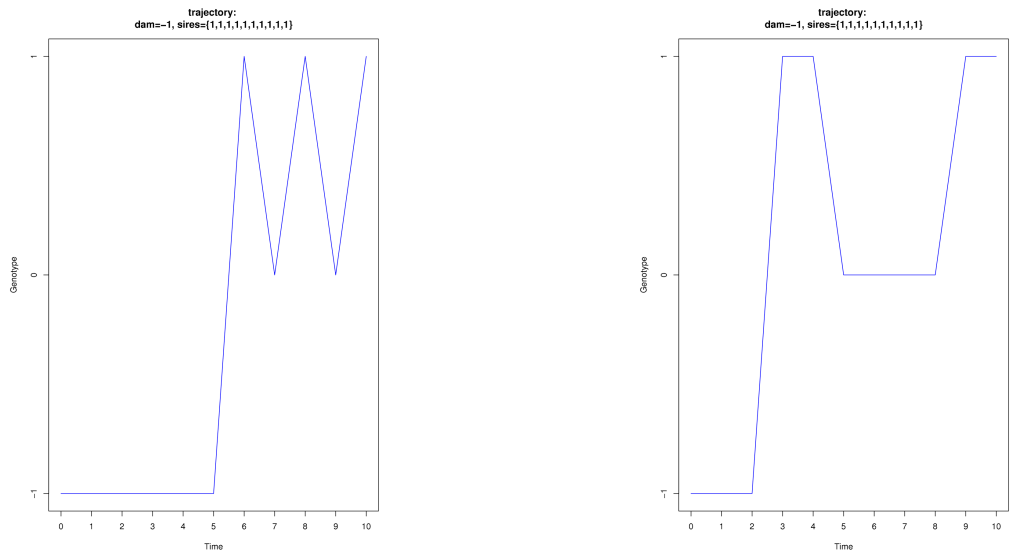}
\end{center}
\caption{ {\bf Two randomly chosen genotype trajectories. The
dependent variable is a genotype value at the fixed index slot of
the array of main herd, see in text for details. } }
\label{fig:1ab}
\end{figure}

\begin{figure}[h]
\begin{center}
\includegraphics[width=3in]{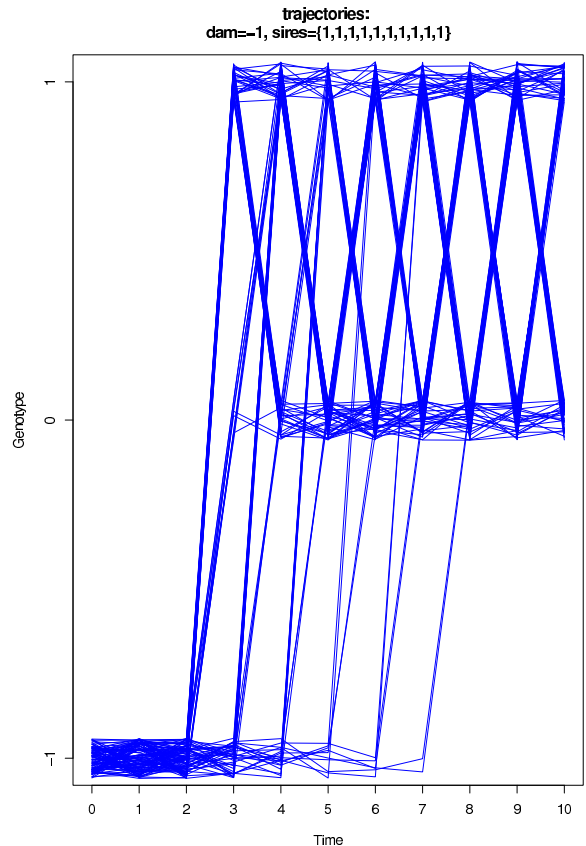}
\end{center}
\caption{ {\bf Bundle of 100 genotype trajectories.} }
\label{fig:2}
\end{figure}

\begin{figure}[h]
\begin{center}
\includegraphics[width=5.5in]{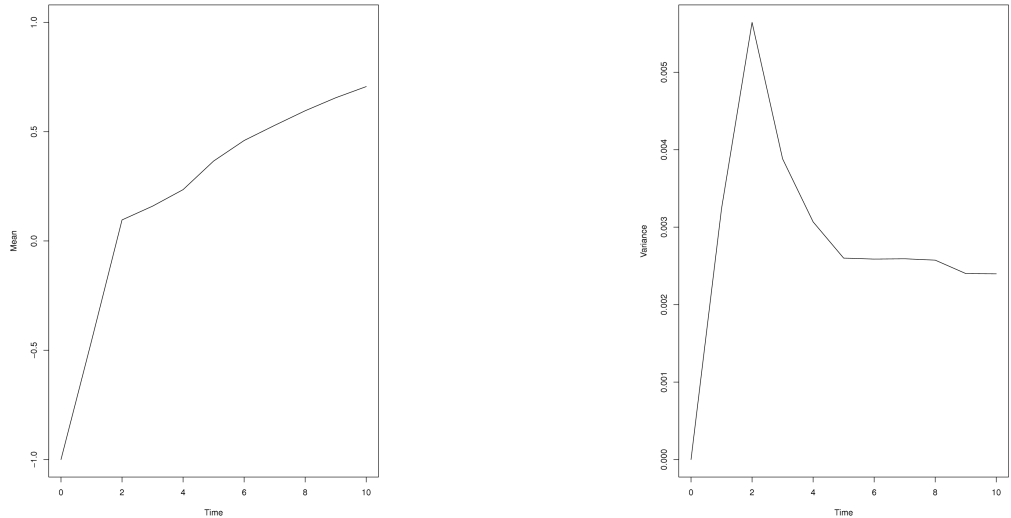}
\end{center}
\caption{ {\bf Mean (left) and variance of mean (right) of
genotype value for 1000 herds, every main herd  size is 200,
replacement herd size is 100 at initial time, $\lambda_D=0.114,
\lambda_d=0.25$.} } \label{fig:4ab}
\end{figure}

In Figure~\ref{fig:1ab} we demonstrate two realizations of single
trajectory in this case.
It  {is} worth mentioning that every plot shows the dependence of
gene index in {a particular, say the j-th,} slot in the array of
animals in the herd. Then in Figures~\ref{fig:1ab}a,b only jumps
can be definitely interpreted as animal change, whereas horizontal
lines could correspond, in principle, replacement at any time step
an animal by
another one with the same genotype. 
 We stress that single trajectories ( Figure~\ref{fig:1ab}) and bundles of trajectories ( Figure~\ref{fig:2}) for the transition of an SNP from one state into another may differ quite markedly from the mean ( Figure~\ref{fig:4ab}(a)).

Next, in Figure~\ref{fig:2}, we  plot a bundle of 100 trajectories
with jittering so that the probability of each route may be
inferred from the plot. {The density of lines allows visual
estimation of the number of animals in particular states
{-1,0,1}.}

\begin{figure}[h]
\begin{center}
\includegraphics[width=3in]{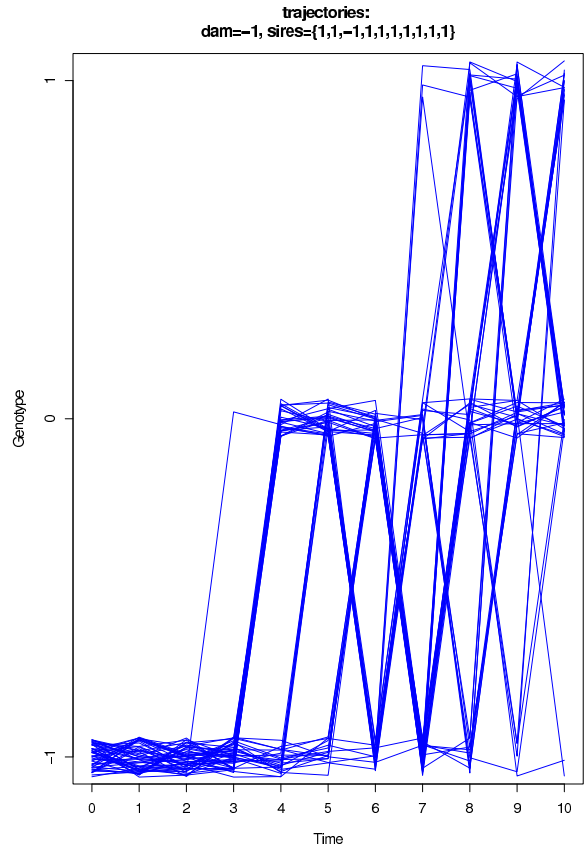}
\end{center}
\caption{ {\bf Bundle of 50 genotype trajectories, one bull in
sequence (third one) with $g=-1$.} } \label{fig:3}
\end{figure}

Finally, in Figure~\ref{fig:3}, we illustrate the effect when one
of the sires has the genotype $-1$ and {again} consider 100
individual trajectories. As one can see from this plot, a single
fault (using a sire with $g=-1$), can seriously slow down the
transition period. It needs to be pointed out that in
Figures~\ref{fig:2}~and~\ref{fig:3} the near vanishing of genotype
-1 at the end of the simulation period does not mean that it has
been eliminated from the whole of the herd.

{Now we turn to the statistical characteristics of ensembles of
simulated herds.} In Figure~\ref{fig:4ab} we show the temporal
evolution of mean (a) and variance (b) for the case of 1000 herds,
{each simulation} {starting} from a ``herd'' of 200 homozygous
dams in the main herd (all initial genotypes are $-1$)  and {then}
developing independently.

\begin{figure}[h]
\begin{center}
\includegraphics[width=3in]{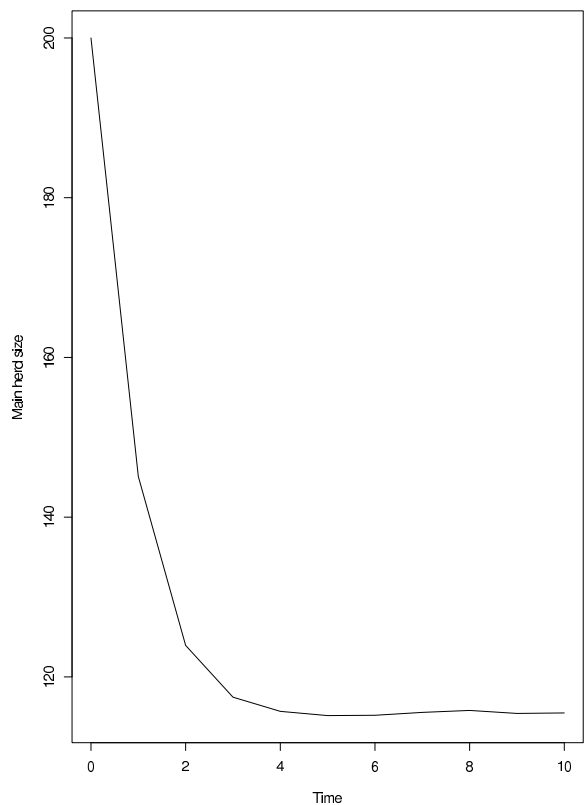}
\end{center}
\caption{ {\bf The dependence of the mean main herd size upon
time. All parameters are the same as in Figure~\ref{fig:4ab}. } }
\label{fig:5}
\end{figure}

An important feature discovered in this simulation is that the
herd size evolves in time in direction of stabilization at a level
which is less than what was chosen as a upper bound in simulation.
The appropriate plot for herd size mean value  is shown in
Figure~\ref{fig:5}. It looks like some sort of ``partial
extinction'' and can be explained as follows. Any positive
fluctuation in number of female born at definite time step is cut
by application of the rule of the upper bound control policy of
assumptions 1 and 2.

In contrast to this,  rare strong negative  fluctuations in
number of females born, which produce deficit of cows in the
replacement herd seriously influence subsequent dynamics and lead
to a period of slow herd size restoration. Averaging over herds
{does} not improve the results, as this would include more and
more rare but strong fluctuations.

As we see from the last plot, the dynamics of {the} replacement
herd strongly influence on those of the main herd. In accord with
this, the presented results  on main herd should be considered as
demonstrating  only tendencies, because we did not seek to
investigate the influence of replacement herd management nor try
to optimize it somehow.

\begin{figure}[h]
\begin{center}
\includegraphics[width=4in]{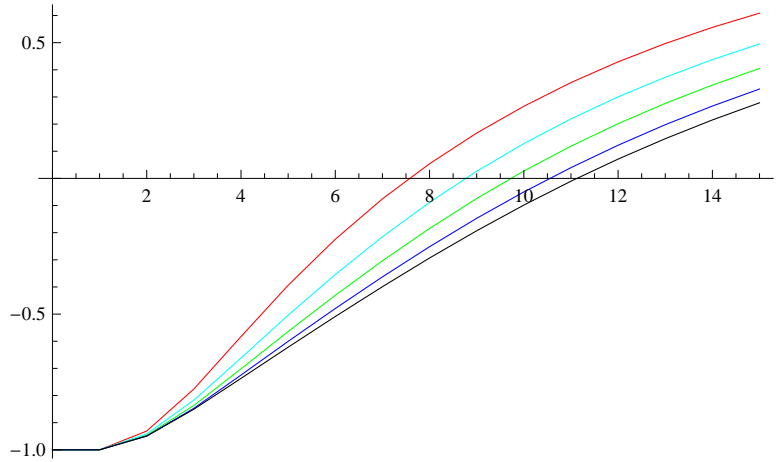}
\end{center}
\caption{ {\bf Mean  genotype value time evolution for 10000
herds, every main herd  size is 400, replacement herd size is 200
at initial time. Colors sequence corresponds to different values
of the mean life constant in the main herd, see details in text.}
} \label{fig:mean-age}
\end{figure}

\begin{figure}[p]
\begin{center}
\includegraphics[width=4in]{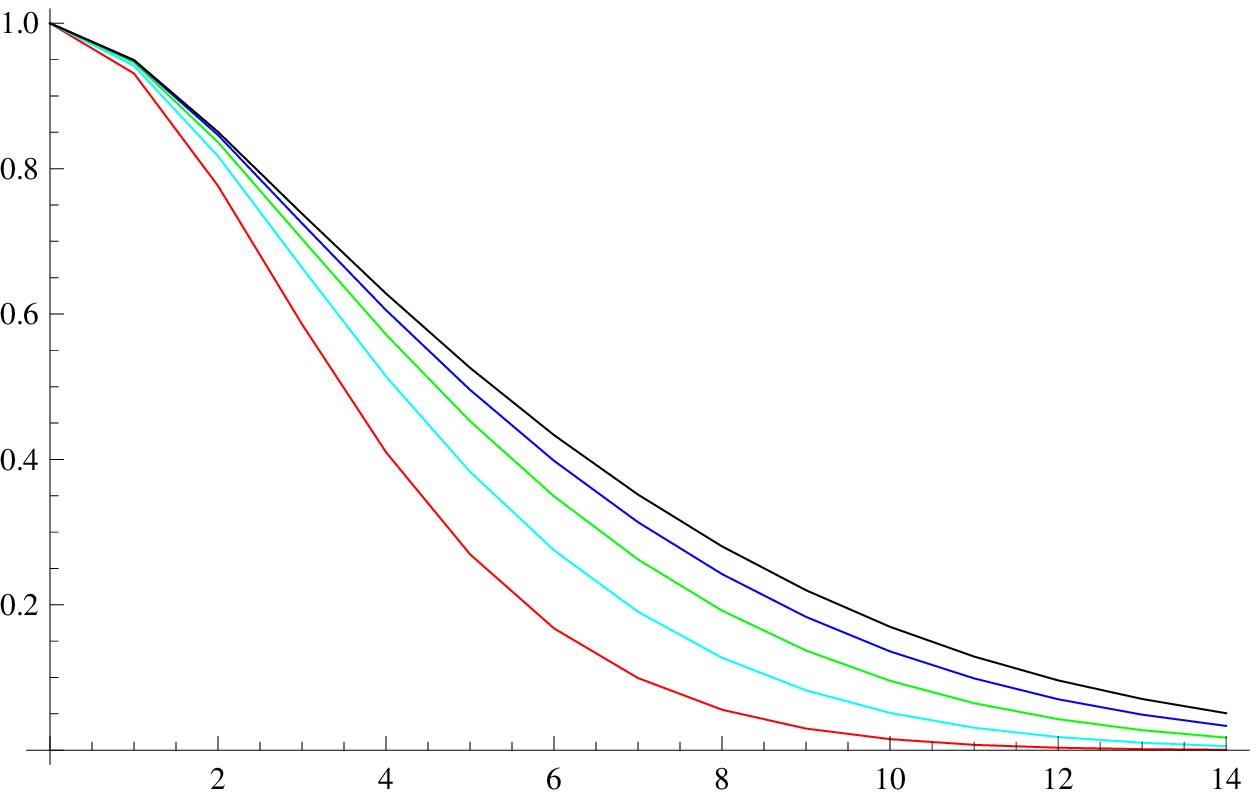}
\end{center}
\caption{ {\bf Mean displayed allele frequency time evolution for
10000 herds, every main herd  size is 400, replacement herd size
is 200 at initial time. Colors sequence corresponds to different
values of the mean life constant in the main herd, see details in
text.} } \label{fig:mean-freq-age}
\end{figure}

\begin{figure}[p]
\begin{center}
\includegraphics[width=4in]{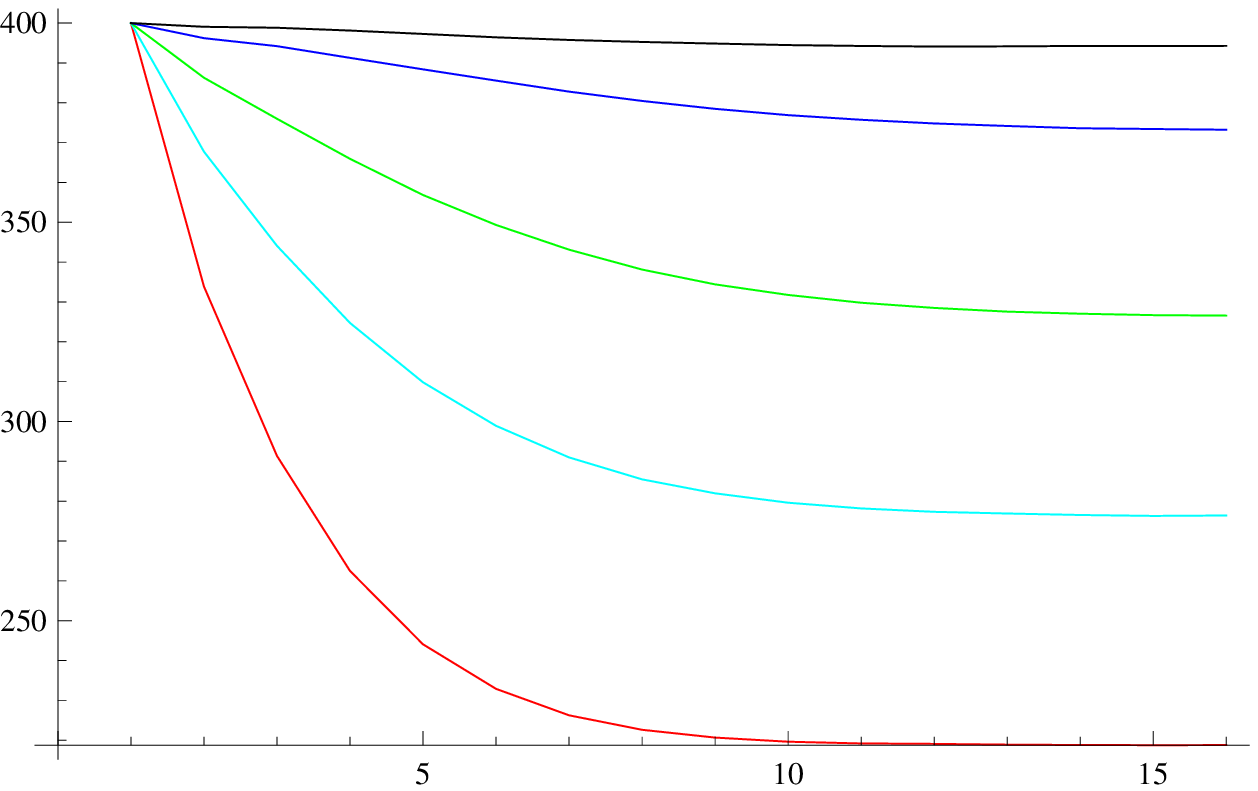}
\end{center}
\caption{ {\bf Mean main herd size time evolution for 10000 herds,
every main herd  size is 400, replacement herd size is 200 at
initial time. Colors sequence corresponds to different values of
the mean life constant in the main herd, see details in text.} }
\label{fig:size-age}
\end{figure}

\begin{figure}[p]
\begin{center}
\includegraphics[width=4in]{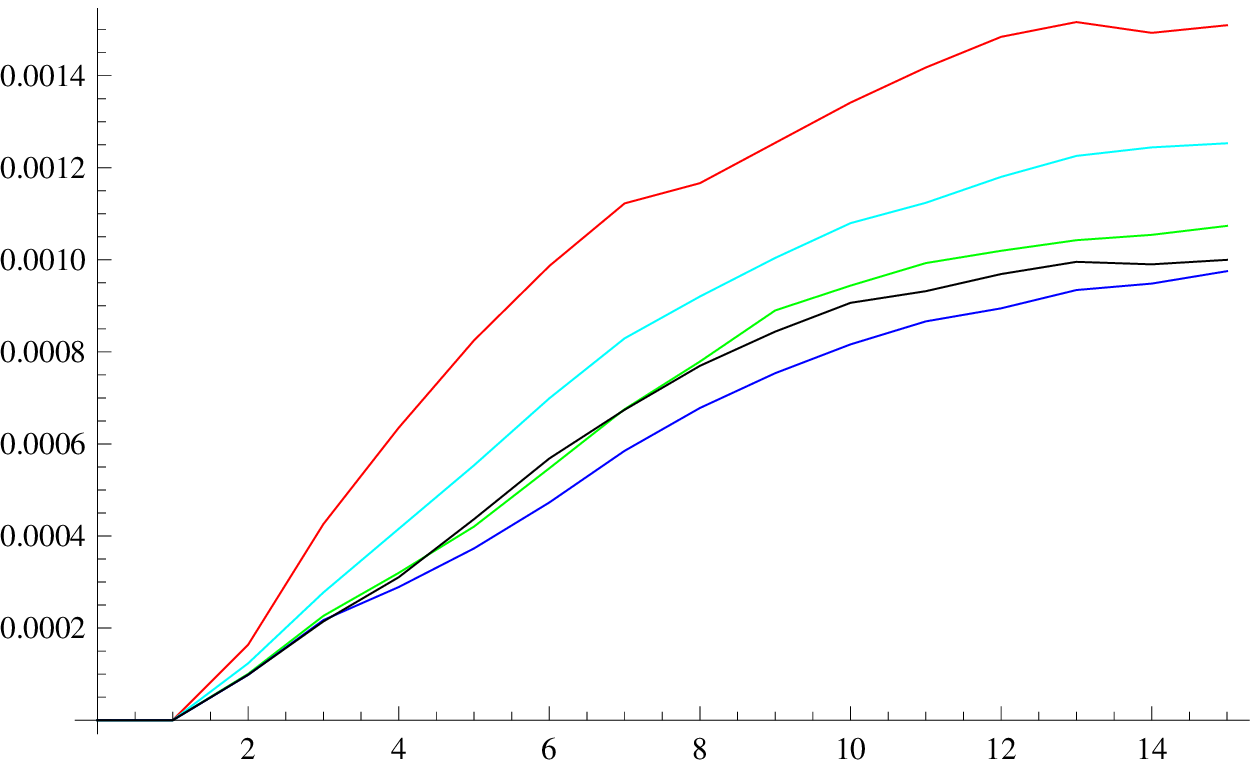}
\end{center}
\caption{ {\bf Variance of mean (right) of genotype value
evolution for 10000 herds, every main herd  size is 400,
replacement herd size is 200 at initial time. Colors sequence
corresponds to different values of the mean life constant in the
main herd, see details in text.} } \label{fig:var-age}
\end{figure}

\begin{figure}[p]
\begin{center}
\includegraphics[width=4in]{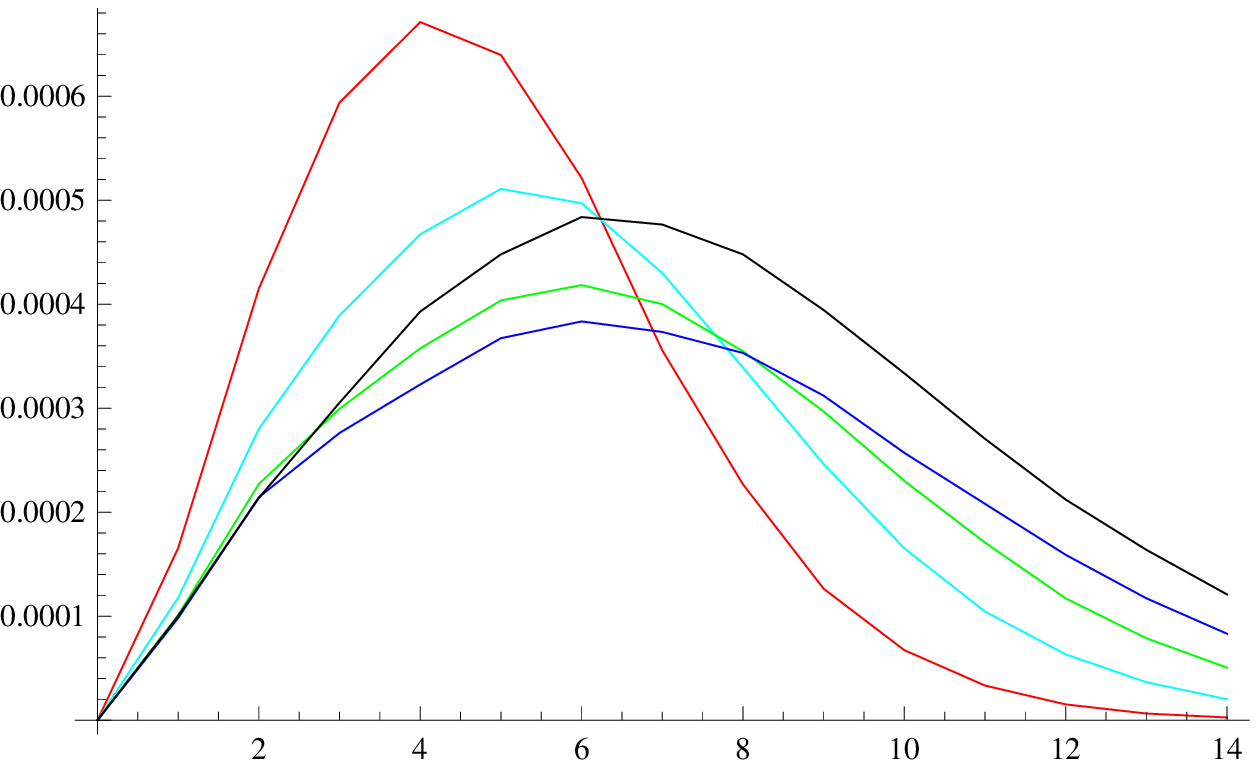}
\end{center}
\caption{ {\bf Variance of the displaced allele frequency for
10000 herds, every main herd  size is 400, replacement herd size
is 200 at initial time. Colors sequence corresponds to different
values of the mean life constant in the main herd, see details in
text.} } \label{fig:var-freq-age}
\end{figure}

As we have already mentioned the at first sight an unpleasant
``extinction effect'' can be eliminated by animal inflow from the
market in a way been discussed above. But much more profitable
seems to be the following point of view, namely that ``partial
extinction'' is a key ingredient in faster switch to a given
allele. In fact, the Poisson process constant $\lambda_D$ for the
main herd is directly linked with ``partial extinction'' level as
one can see from the following plots in Figures~
\ref{fig:mean-age}, \ref{fig:mean-freq-age}, \ref{fig:size-age},
\ref{fig:var-age}, \ref{fig:var-freq-age} where we demonstrate the
rate of the switch from a definite sign (it was chosen as $-$) of
the allele into opposite sign depended on values of $\lambda_D$.
Values $\lambda_D$ and colors sequence (red, cyan, green, blue,
black) corresponds to probability equals 0.8 for a cow to live in
herd up to $4,6,8,10,12$ years correspondingly, it was used
averaging over 10000 herds in this simulations.

\begin{figure}[p]
\begin{center}
\includegraphics[width=4in]{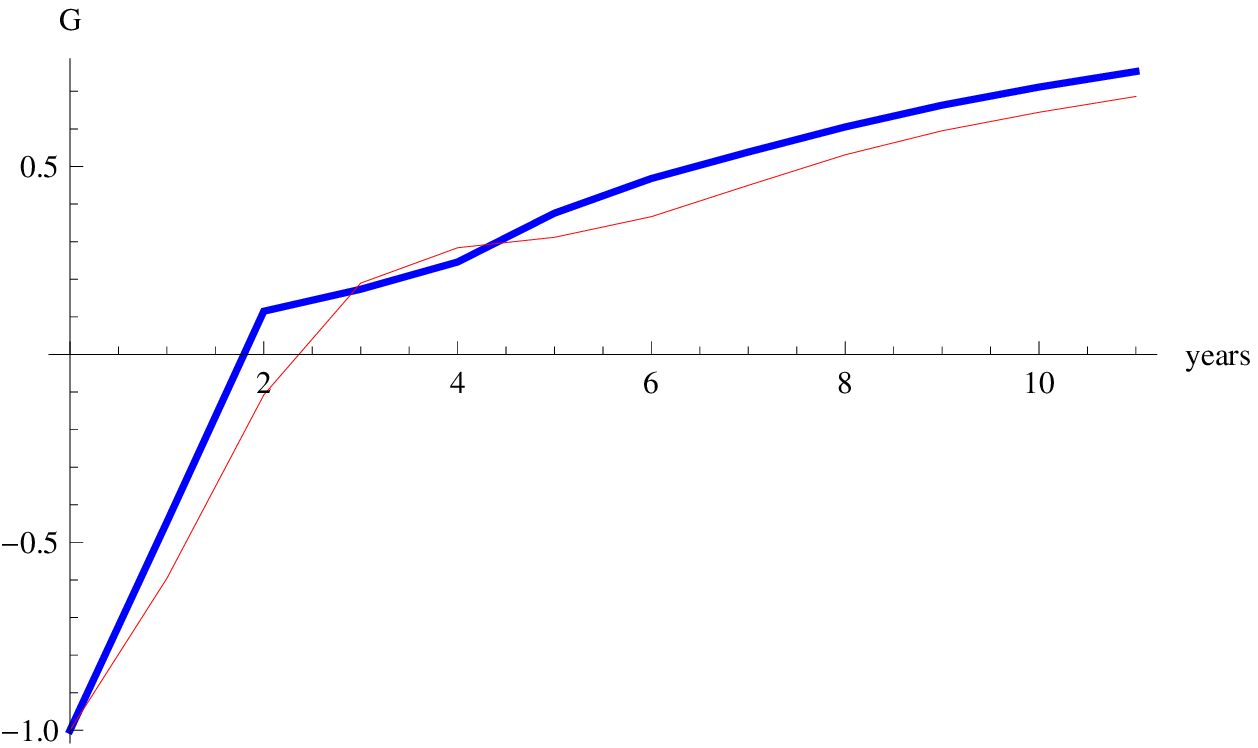}
\end{center}
\caption{ {\bf The comparison of the mean main herd gene index
upon time dependencies   for two models. Thin red line is for
model 1, thick-blue line is for model 2. All parameters are the
same as in Figure~\ref{fig:4ab}. } } \label{fig:2models-mean}
\end{figure}

\begin{figure}[p]
\begin{center}
\includegraphics[width=4in]{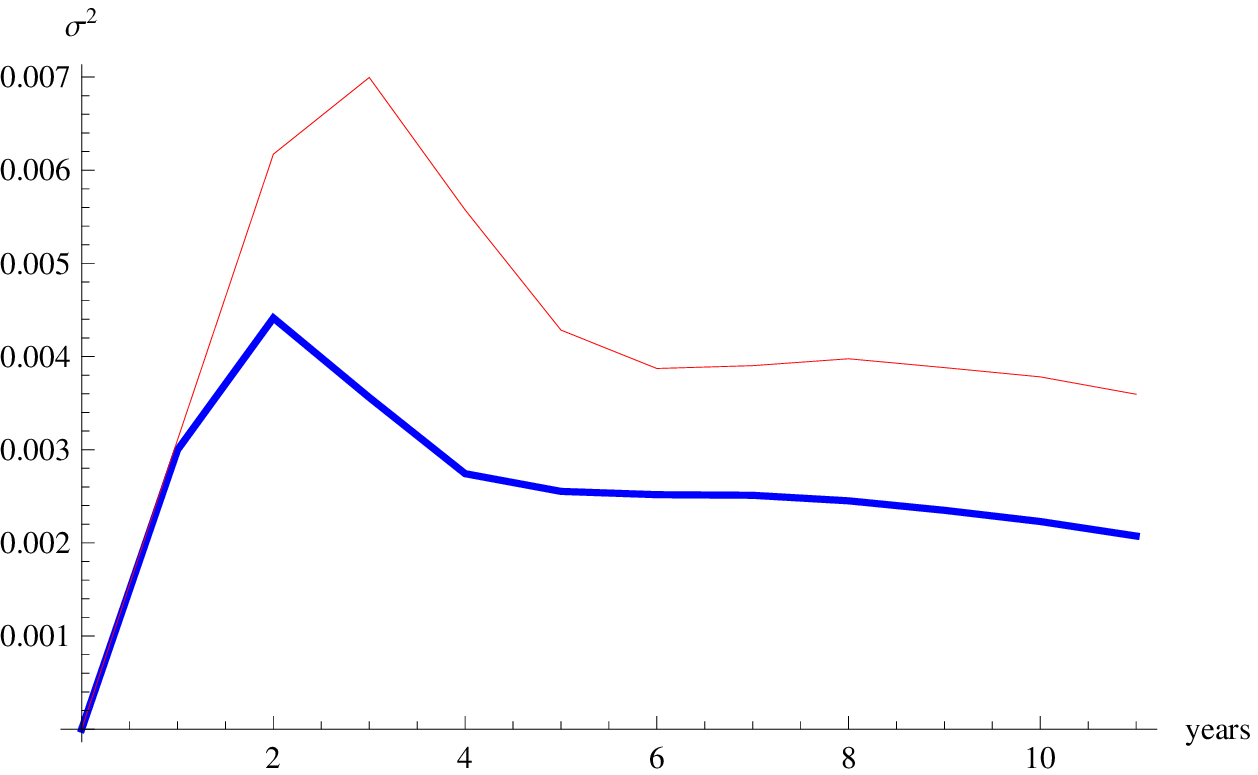}
\end{center}
\caption{ {\bf The comparison of the mean main herd gene index
variance upon time dependencies for two models. Thin red line is
for model 1, thick-blue line is for model 2. All parameters are
the same as in Figure~\ref{fig:4ab}. } } \label{fig:2models-var}
\end{figure}

\begin{figure}[htb]
\begin{center}
\includegraphics[width=4in]{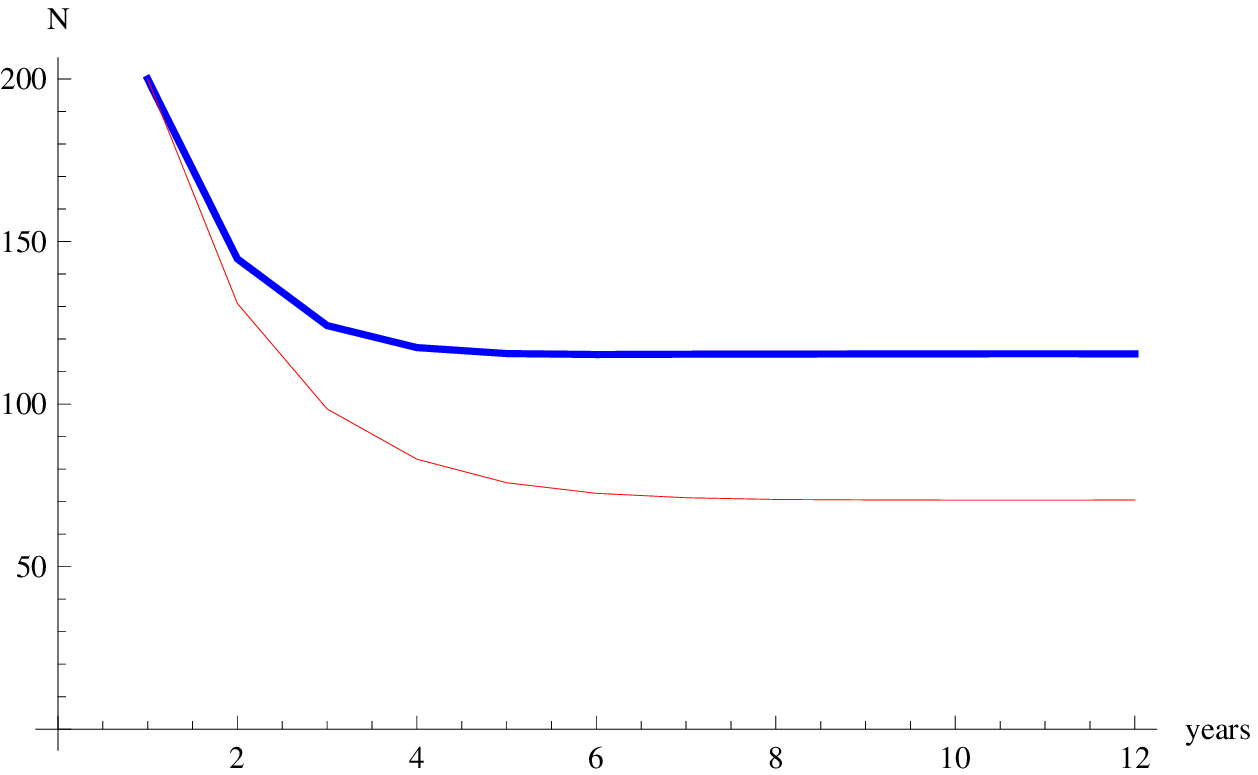}
\end{center}
\caption{ {\bf The comparison of the mean main herd size upon time
dependencies for two models. Thin red line is for model 1,
thick-blue line is for model 2. All parameters are the same as in
Figure~\ref{fig:4ab}. } } \label{fig:2models-size}
\end{figure}

One of the interesting feature of the model is the sensitivity to
a fine detail of the transition of animal from  replacement herd
to a main one. We consider two variants, in the first one we fill
the main herd up to a limit size first from the replacement herd,
then make a random choice of a way for rest animals in replacement
hers to leave the herd. In the second variant, we make a random
choice of leave/rest and then specify the variant of leave (if
success). It turns out that at intermediate time these two
slightly different procedure give observably different behaviour,
as demonstrated in Figure~\ref{fig:2models-mean} for mean values,
variances Figure~\ref{fig:2models-var} and mean size of the main
herd Figure~\ref{fig:2models-size}. For gene index dynamics both
curves are very near, variances differ slightly but the mean herd
size is influence strongly by the control scheme. The last leads
to conclusion that it is necessary to be very accurate when
formulating any control scheme for such a dynamical system,
schemes seems to be very near at first glance could produce
significantly different results.

\section*{Discussion}

 To summarize, we have constructed a system of stochastic differential equation that can model temporal evolution of biallelic polymorphism in a  deme under conditions of controlled mating. The model incorporates peculiarities typical for New Zealand dairy herd management such as herd split into main and replacement herds, typical lifetime distribution, size control for main herd and rule of inflow from the replacement into main herd. Currently the model is implemented in R but an open source C++ version is under development \cite{Guy}.

 Simulations have demonstrated that when a maximum herd size is imposed local fluctuations of new born animals will strongly influence the system dynamics and lead to observable diminishing of the herd size (partial extinction). To
 suppress this feature, which is not observed in real farm situations, the  model has been  further adjusted to allow for an external inflow of animals from the market. Another important conclusion one can make is that the investigation of replacement herd  management policy could be of great importance for reaching optimization goals.

\section*{Acknowledgments}
Our thanks go to Jack Hooper of Livestock Improvement Corporation
for reading drafts of this article and for valuable comments on
existing herd management practice in New Zealand.


\newpage

\begin{table}[h]
\caption{ \bf{Hardy-Weinberg Conditions and the New Zealand Dairy
Herd}}
\begin{tabular}{|l|l|}
  \hline
  Ideal Population & Production industry \\
  \hline\hline
  Large homogeneous population & Heterogeneous Herds and Regions \\ \hline
  Random mating & Controlled mating \\ \hline
  No mutation & Mutations happen \\ \hline
  No selection & Selection exists \\ \hline
  No migration & Export/import of genes \\
  \hline
\end{tabular}
\label{hwtab}
\end{table}

\end{document}